\def\presentation{
\voffset  0.2 truecm 
\hoffset -.19in
\oddsidemargin 0in \evensidemargin 0in
\marginparwidth .75in \marginparsep 7pt \topmargin 0in
\headheight 12pt \headsep .25in
\footheight 18pt \footskip .35in
\textheight 21.0 truecm \textwidth 16.0 truecm 
\columnsep 10pt \columnseprule 0pt }
\documentstyle{article}
\presentation
\begin{document}
%

%%%%%%%%%%%%%%%%%%%%%%%%%%%%%%%%%%%%%%%%%%%%%%%%%%%%%%%%%%%%%%%%%%%%%
%		DEFINITIONS FOR TEX
%%%%%%%%%%%%%%%%%%%%%%%%%%%%%%%%%%%%%%%%%%%%%%%%%%%%%%%%%%%%%%
%
\def\tilde{\widetilde}
\def\bar{\overline}
\def\hat{\widehat}
\def\*{\star}
\def\[{\left[}
\def\]{\right]}
\def\({\left(}
\def\){\right)}
\def\zb{{\bar{z} }}
\def\frac#1#2{{#1 \over #2}}
\def\inv#1{{1 \over #1}}
\def\half{{1 \over 2}}
\def\d{\partial}
\def\der#1{{\partial \over \partial #1}}
\def\dd#1#2{{\partial #1 \over \partial #2}}
\def\vev#1{\langle #1 \rangle}
\def\bra#1{{\langle #1 |  }}
\def\ket#1{ | #1 \rangle}
\def\rvac{\hbox{$\vert 0\rangle$}}
\def\lvac{\hbox{$\langle 0 \vert $}}
\def\2pi{\hbox{$2\pi i$}}
\def\e#1{{\rm e}^{^{\textstyle #1}}}
\def\grad#1{\,\nabla\!_{{#1}}\,}
\def\dsl{\raise.15ex\hbox{/}\kern-.57em\partial}
\def\Dsl{\,\raise.15ex\hbox{/}\mkern-.13.5mu D}
\def\comm#1#2{ \BBL\ #1\ ,\ #2 \BBR }
\def\x{\stackrel{\otimes}{,}}
\def\det{ {\rm det}}
\def\tr{{\rm tr}}
%
%%%%%%%%%%%%%%%%%%%%GREEK LETTERS%%%%%%%%%%%%%%%%%%%%%%%%%%%%%%
%
\def\th{\theta}		\def\Th{\Theta}
\def\ga{\gamma}		\def\Ga{\Gamma}
\def\be{\beta}
\def\al{\alpha}
\def\ep{\epsilon}
\def\la{\lambda}	\def\La{\Lambda}
\def\de{\delta}		\def\De{\Delta}
\def\om{\omega}		\def\Om{\Omega}
\def\sig{\sigma}	\def\Sig{\Sigma}
\def\vphi{\varphi}
%
%%%%%%%%%%%%%%%%%%%CALIGRAPHIC LETTERS%%%%%%%%%%%%%%%%%%%%%%%%%
%
\def\CA{{\cal A}}	\def\CB{{\cal B}}	\def\CC{{\cal C}}
\def\CD{{\cal D}}	\def\CE{{\cal E}}	\def\CF{{\cal F}}
\def\CG{{\cal G}}	\def\CH{{\cal H}}	\def\CI{{\cal J}}
\def\CJ{{\cal J}}	\def\CK{{\cal K}}	\def\CL{{\cal L}}
\def\CM{{\cal M}}	\def\CN{{\cal N}}	\def\CO{{\cal O}}
\def\CP{{\cal P}}	\def\CQ{{\cal Q}}	\def\CR{{\cal R}}
\def\CS{{\cal S}}	\def\CT{{\cal T}}	\def\CU{{\cal U}}
\def\CV{{\cal V}}	\def\CW{{\cal W}}	\def\CX{{\cal X}}
\def\CY{{\cal Y}}	\def\CZ{{\cal Z}}
%
%%%%%%%%%%%%%%% MATH CHARACTERS %%%%%%%%%%%%%%%%%%%%%%%%%%%%
%
\font\numbers=cmss12
%\font\numbers=cmu10 scaled\magstep1
\font\upright=cmu10 scaled\magstep1
\def\stroke{\vrule height8pt width0.4pt depth-0.1pt}
\def\topfleck{\vrule height8pt width0.5pt depth-5.9pt}
\def\botfleck{\vrule height2pt width0.5pt depth0.1pt}
\def\Zmath{\vcenter{\hbox{\numbers\rlap{\rlap{Z}\kern
		0.8pt\topfleck}\kern
		2.2pt \rlap Z\kern 6pt\botfleck\kern 1pt}}}
\def\Qmath{\vcenter{\hbox{\upright\rlap{\rlap{Q}\kern
                   3.8pt\stroke}\phantom{Q}}}}
\def\Nmath{\vcenter{\hbox{\upright\rlap{I}\kern 1.7pt N}}}
\def\Cmath{\vcenter{\hbox{\upright\rlap{\rlap{C}\kern
                   3.8pt\stroke}\phantom{C}}}}
\def\Rmath{\vcenter{\hbox{\upright\rlap{I}\kern 1.7pt R}}}
\def\Z{\ifmmode\Zmath\else$\Zmath$\fi}
\def\Q{\ifmmode\Qmath\else$\Qmath$\fi}
\def\N{\ifmmode\Nmath\else$\Nmath$\fi}
\def\C{\ifmmode\Cmath\else$\Cmath$\fi}
\def\R{\ifmmode\Rmath\else$\Rmath$\fi}
%%%%%%%%%%%%%%%%%%%%%%%%%%%%%%%%%%%%%%%%%%%%%%%%%%%%%%%%%%%%%%%%%
\def\cadremath#1{\vbox{\hrule\hbox{\vrule\kern8pt\vbox{\kern8pt
			\hbox{$\displaystyle #1$}\kern8pt} 
			\kern8pt\vrule}\hrule}}
\def\proof{\noindent Proof. \hfill \break}
\def\cqfd{ {\hfill{$\Box$}} }
\def\square{\hfill
\vrule height6pt width6pt depth1pt \\}
%
%%%%%%%%%%%%%%%%%%% LATEX SPECILIALITIES %%%%%%%%%%%%%%%%%%%%%%%
%
\def\debut{ \begin{eqnarray} }
\def\fin{ \end{eqnarray} }
\def\non{ \nonumber }
%
%%%%%%%%%%%%%%%%%%%%%%%%%%%%%%%%%%%%%%%%%%%%%%%%%%%%%%%%%%%%%%%%%
%		 END OF DEFINITIONS 
%%%%%%%%%%%%%%%%%%%%%%%%%%%%%%%%%%%%%%%%%%%%%%%%%%%%%%%%%%%%%%%%
%
%%%%%%%%%%%%%%%%%%%%%%%%%%%%%%%%%%%%%%%%%%%%%%%%%%%%%%%%%%%%%%%%%
%
\rightline{SPhT-96-138, LPTHE-96-53}
  ~\vskip 1cm
\centerline{{\LARGE Form factors, KdV and Deformed Hyperelliptic Curves.}
\footnote[9]{for the proceedings of the conference ``Advanced Quantum
Field Theory (in memory of Claude Itzykson)", Lalonde,
France, Sept. 96.}  }
\vskip 1cm
\centerline{\large O. Babelon ${}^{a\ 0}$, D. Bernard ${}^b$
\footnote[0]{Membre du CNRS} and F.A. Smirnov ${}^a$
\footnote[1]{On leave from Steklov Mathematical Institute,
Fontanka 27, St. Petersburg, 191011, Russia} }
\vskip1cm
\noindent  ${}^a$ Laboratoire de Physique Th\'eorique et Hautes
Energies \footnote[2]{\it Laboratoire associ\'e au CNRS.}\\
 ~~~ Universit\'e Pierre et Marie Curie, 
4 place Jussieu, 75252 Paris cedex 05-France.\\
\bigskip 
${}^b$ Service de Physique Th\'eorique de Saclay
\footnote[3]{\it Laboratoire de la Direction des Sciences de la
Mati\`ere du Commissariat \`a l'Energie Atomique.},
F-91191, Gif-sur-Yvette, France.\\
 \vskip 0.5 truecm
\rightline{\`A la m\'emoire de Claude ITZYKSON.}
 \vskip 0.5 truecm 
%\vfill
%\newpage
%
We review and summarize recent works on the relation
between form factors in integrable quantum field theory
and deformation of geometrical data associated to
hyper-elliptic curves. This relation, which is based on a 
deformation of the Riemann bilinear identity, in particular 
leads to the notion of null vectors in integrable field 
theory and to a new description of the KdV hierarchy.
\def\b{ \beta }
%
%
% 
%%%% DEBUT  %%%%%%%%%%%%%%%%%%%%%%%%%
%
%\section{Introduction.}
%
%
\section{Form factor formula.}
Let us first recall what form factors are.
We shall consider the Sine-Gordon theory.
The Sine-Gordon equation follows from the action:
$$ S={\pi\over\gamma}\int \CL \ d^2 x,\qquad with\quad
\CL=(\partial _{\mu}\varphi)^2+
m^2 (\cos (2\varphi)-1)  $$
where $\gamma$ is the coupling constant, $0<\gamma<\pi$.
The free fermion point is at $\ga=\frac{\pi}{2}$.
In the quantum theory, the relevant coupling constant is:
$$\xi ={\pi \gamma\over \pi -\gamma}. $$
We shall always use the constant $\xi$, which plays the role
of the Planck constant.

We shall actually not consider the Sine-Gordon model but a restriction of it.
The Sine-Gordon theory contains two subalgebras of local operators
which, as operator algebras are generated by $\exp (i\varphi)$
and $\exp (-i\varphi)$ respectively. Let us concentrate on one
of them, say the one generated by $\exp (i\varphi)$. It is known that
this subalgebra can be considered independently of the rest of the
operators as the operator algebra of the theory with 
a modified energy-momentum tensor.
This modification changes the trace of the stress tensor, and therefore
changes the ultraviolet limit of the correlation functions. 
This modification corresponds to the restricted Sine-Gordon theory (RSG).
For rational ${\xi\over\pi}$ it describes
the $\Phi _{1,3}$-perturbations of the minimal models of CFT,
but it can be considered for generic values of $\xi$ as well.

The asymptotic states of the Sine-Gordon theory is made of 
solitons, anti-solitons and their bound states. We will
denote $n$ solitons, $n$ anti-solitons states by:
\debut
|\b _1,\b _2,\cdots ,\b _{2n}\rangle\rangle _{\epsilon _{1}, \epsilon _{2}
\cdots ,\epsilon _{2n}} \non
\fin
We shall consider the case $\xi ={\pi\over \nu}$ for
$\nu =1, 2,\cdots$, when the reflection of solitons and
anti-solitons is absent. 
The S-matrix is then diagonal and given by
$$S(\b)=
%\prod\limits _{j=1}^{\nu -1}{\sinh{1\over 2}(\b +{\pi i\over\nu}j)
%\over\sinh{1\over 2}(\b -{\pi i\over\nu}j)}= 
\prod_{j=1}^{\nu -1} \left( {B q^j -1 \over B -q^j } \right),
\quad with \quad q=e^{i\frac{\pi}{\nu}}$$
We shall use the following notations:
$B=\exp (\b)$ and $ b=\exp ( {2\pi \over\xi} \b)=\exp (2\nu \b)$

The form factors are the matrix elements of local fields between
two asymptotic states. By crossing symmetry they can
be computed from the form factors between the vaccum  and
any $n$ solitons, $n$ anti-solitons states~:
\debut
f_{\cal O}(\b _1,\b _2,\cdots ,\b _{2n})_{\epsilon _{1}, \epsilon _{2}
\cdots ,\epsilon _{2n}}=
\langle\langle 0|\CO (0)
|\b _1,\b _2,\cdots ,\b _{2n}\rangle\rangle _{\epsilon _{1}, \epsilon _{2}
\cdots ,\epsilon _{2n}} \non
\fin
where $\CO(x)$ denotes any local operator.
The next section is devoted to a brief description of 
integral formula for these form factors.

\subsection{Form factors at the reflectionless points.}
At the reflectionless points ($\xi={\pi\over\nu}$, $\nu=1,2,\cdots$)
there is a wide class of local operators $\CO$ for which
the form factors in the (restricted) Sine-Gordon model
corresponding to a state with $n$-solitons and
$n$-anti-solitons are given by
\def\b{\beta}
\debut
 &&~~~~~~~~~~~~~~~~~
f_\CO (\b _1,\b _2,\cdots ,\b _{2n})_{- \cdots -+\cdots +}=
\label{ff} \\
&=&c^n e^{-{1\over 2}(\nu (n-1)-n)\sum_j \b _j}~
\prod\limits _{i<j}\zeta (\b _i-\b _j)
\prod\limits _{i=1}^n \prod\limits _{j=n+1} ^{2n}
{1\over\sinh \nu (\b _j-\b _i -\pi i)}
\ \widehat{f}_\CO (\b _1,\b _2,\cdots ,\b _{2n})_{-\cdots -+\cdots +}
\non
\fin
The function $\zeta (\b )$, without poles in the
strip $0< Im ~ \beta ~< 2\pi$, satisfies $\zeta (-\beta) =
S(\beta) \zeta (\beta)$ and $\zeta (\beta-2\pi i) = \zeta (-\beta)$.
The S-matrix $S(\beta)$ is defined above.
The constant $c$ is given by $c=2\nu(\zeta(-i\pi))^{-1}$.
The most essential part of the form factor is given by \cite{book}:
\debut
\widehat{f}_\CO (\b _1,\b _2,\cdots ,\b _{2n})_{-\cdots -+\cdots +}&=&
\label{ints}\\
&&\hskip -5cm
={1\over (2\pi i)^n}\int dA_1\cdots \int dA_n
\prod\limits _{i=1}^n \prod\limits _{j=1} ^{2n} \psi (A_i,B_j)
\prod\limits _{i<j} (A_i^2-A_j^2)
\ L_\CO^{(n)} (A_1,\cdots ,A_n|B_1,\cdots ,B_{2n})
\prod\limits _{i=1}^n a_i^{-i} \non
\fin
where $B_j=e^{\b _j}$ and
\debut
 \psi (A,B)=\prod\limits _{j=1}^{\nu -1}(B -Aq^{-j} ),
\qquad {\rm with}~~ q= e^{i\pi/\nu}
\label{defpsi}
\fin
As usual we define $a=A^{2\nu}$.
Here and later if the range of integration is not specified the
integral is taken around 0. Notice that the operator dependence
of the form factors (\ref{ff}) only enters in $\widehat{f}_\CO$.
 
Different local operators ${\cal O}$ are defined by different
functions $L_\CO^{(n)} (A_1,\cdots ,A_n|B_1,\cdots ,B_{2n}) $.
These functions are symmetric polynomials of $A_1,\cdots ,A_n$.
For the primary operators $\Phi _{2k}=\exp (2k i\varphi )$
and their Virasoro descendants, $L_\CO$ are symmetric Laurent
polynomials of $ B_1,\cdots ,B_{2n} $. For the
primary operators $\Phi _{2k+1}=\exp ((2k+1) i\varphi)$,
they are symmetric Laurent polynomials of $ B_1,\cdots ,B_{2n} $
multiplied by $\prod B_j ^{1\over 2}$.
Our definition of the fields $\Phi _{m} $ is related to the notations coming
from CFT as follows: $\Phi _{m}$ corresponds to $ \Phi _{[1,m+1]}$.
The requirement of locality for the operator $\CO$
is guaranteed by the following simple
recurrent relation for the polynomials $L_{\CO}^{(n)}$:
\debut
&&\left . L_\CO^{(n)} (A_1,\cdots ,A_n|B_1,\cdots ,B_{2n})
\right|_{B_{2n}=-B_{1},\ A_n=\pm B_{1}}= \non\\
&&\hskip 1cm
=-\epsilon^\pm L_\CO^{(n-1)} (A_1,\cdots ,A_{n-1}|B_2,\cdots ,B_{2n-1})
\label{res}
\fin
where $\epsilon =+$ or $-$ respectively for the operators $\Phi _{2k} $
and their descendents, or for $\Phi _{2k+1} $ and their descendents.
In addition to the simple formula (\ref{ints}) we have to add
the requirement
\debut res _{A_n=\infty}\(
\prod\limits _{i=1}^n \prod\limits _{j=1} ^{2n} \psi (A_i,B_j)
\prod\limits _{i<j} (A_i^2-A_j^2)
\ L_\CO^{(n)} (A_1,\cdots ,A_n|B_1,\cdots ,B_{2n})a _n^{-k}\)=0,
\quad k\ge n+1
\label{vv}\fin
This is true in particular if $deg _{A_n}(L_{\CO})<2\nu$,
and therefore the restriction (\ref{vv})
disappears only in the classical limit $\nu\to\infty$.
This class of local operators is not complete
for the reason that the anzatz (\ref{ff}) is too restrictive.
We obtain the complete set of operators only in
the classical limit. However there is a possibility to
define the form factors of local operators which correspond to
polynomials satisfying the relation (\ref{res}) without any
restriction of the kind (\ref{vv}). To do that for the reflectionless
points one has to consider the coupling constant in generic
position (in which case the formulae for the form factors are
much more complicated \cite{book}) and to perform carefully the limit
$\xi={\pi\over\nu}+\epsilon$, $\epsilon\to 0$.
An example of such calculation
for $\xi =\pi$ is given in \cite{sm1}.
We would like to emphasize that the local operator
can be defined for any polynomial satisfying (\ref{res}) but
its form factors are not necessarily given by the anzatz (\ref{ff}).
Physically the existence of local operators for the
reflectionless case whose form factors are not given by the
anzatz (\ref{ff}) is related to the existence of additional
local conserved quantities which constitute the algebra $\widehat{sl}(2)$.
In spite of the fact that the form factors of the form (\ref{ff})
do not define all the operators they provide a good example for
explaining the properties valid in generic case.
 
The explicit form of the polynomials $L_\CO$ for the primary operators
$\Phi_m= e^{im\varphi}$ is as follows
$$ L_{\Phi _m}^{(n)}(A_1,\cdots ,A_n|B_1,\cdots ,B_{2n})
=\prod\limits _{i=1}^n A_i^m
\prod\limits _{j=1}^{2n} B_j^{-{m\over 2}} $$
We shall consider the Virasoro descendents of the
primary fields. We shall restrict ourselves by considering only
one chirality. Obviously, the locality relation (\ref{res}) is not
destroyed if we multiply the polynomial $L_\CO^{(n)}(A|B)$ either by
$I_{2k-1}(B)$ or by $J_{2k}(A|B)$ with
\debut
I_{2k-1}(B)&=&\({1+q^{2k-1}\over1-q^{2k-1}}\)
s_{2k-1}(B),\qquad k=1,2,\cdots\label{defI}\\
J_{2k}(A|B)&=&s_{2k}(A)- {1\over 2}s_{2k}(B),
\qquad k=1,2,\cdots
\label{defJ}
\fin
Here we use the following definition:
$s_k(x_1,\cdots ,x_m)=\sum\limits _{j=1}^{m}x_j^{k}$.

The multiplication by $I_{2k-1}$ corresponds to the application of
the local integrals of motion. The normalization factor
$\({{1+q^{2k-1}\over1-q^{2k-1}}}\) $ is introduced for 
convenience. Since the boost operator acts by dilatation
on $A$ and $B$, $I_{2k-1}$ has spin $(2k-1)$ and $J_{2k}$ has spin $2k$.
 
The crucial assumption which we make is that the space
of local fields descendents of
the operator   $\Phi _{m} $ is generated by
the operators obtained from the generating function
\debut
\CL _m(t,y|A|B)=
\exp\Bigl(\sum\limits _{k\ge 1}t_{2k-1}I_{2k-1}(B)
+y_{2k}J_{2k}(A|B)\Bigr)~~
\({\prod\limits _{i=1}^n A_i^m
\prod\limits _{j=1}^{2n} B_j^{-{m\over 2}} }\) \label{gf1}
\fin
This is our main starting point. As explained below,
this assumption follows from the classical meaning of the variables $A,B$
\cite{bbs}. We will restrict ourselves to the descendents
of the identity operator which correspond to $m=0$ in eq.(\ref{gf1}).
 
\subsection{Form factors and quantization of solitons.}
We now describe how the integral formula can be understood
as arising from a (special) quantization of the quantum
mechanical problem describing the dynamic of a system
of $n$ solitons.

For each $n$-soliton solution
we introduce pairs of conjugated variables $A_i$ and
$P_i$ ($i=1\cdots ,n$),
which in the quantum case  satisfy Weyl commutation relations.
Every local operator $\CO$ can be considered as acting in
this $A$-representation, and therefore can be identified with a
certain operator $\CO (A,P)$. The typical formula for the matrix
element of $\CO$ between two $n$-soliton states can be
presented as \cite{bbs}:
\debut
\vev{B'|\CO |B}=
\int\limits \Psi(A,B')^{\dag}\CO (A,P) \Psi(A,B)~d\mu(A),
\label{premier}
\fin
where $\Psi(A,B)$ is the wave-function of the state of
$n$ solitons with momenta  $B_1,\cdots ,B_n$.
The $P_j$ variables are related to the variables $A_i$ and
$B_k$ by $P_j=\prod_k\({\frac{B_k-A_j}{B_k+A_j}}\)$.
The measure $d\mu (A)$ include a specific weight
admitting a natural interpretation in the $n$-soliton symplectic
geometry.  In formula (\ref{premier}), the variables $A$ are complex.
The integration domain specifies the configuration
space of the quantum mechanical problem.

At the classical level, the conjugated variables $A_i$ and $P_j$ 
arise from a particular paramatrization of the $n$-soliton
solutions of the Sine-Gordon equation. They are naturally related
to the zeroes and poles of the Jost solution of the associated
linear problem. In particular, the Sine-Gordon field can be 
paramatrized in terms of the $A$ and $B$ variables as~:
\debut
e^{i\varphi}= \prod_{j=1}^n\({ \frac{A_j}{B_j}}\) \non
\fin

In comparing the formula (\ref{ints}) and (\ref{premier}), the 
soliton wave functions are more or less identified with the
functions $\prod_{i,j}\psi(A_i,B_j)$ with $\psi(A,B)$ defined in
eq.(\ref{defpsi}), and the integration mesure $d\mu(A)$ in eq.(\ref{premier})
is identified with the Vandermond determinant $\prod_{i<j}(A^2_i-A^2_j)$
in eq.(\ref{ints}). Furtheremore, as explained in \cite{bbs},
the factor $\prod_ia_i^{-i}$
in eq.(\ref{ints}) is related to the positions of the
soliton trajectories in the $A$-plane.

But the most important point is that the polynomials
$L_\CO(A_1,\cdots,A_n|B_1,\cdots,B_n)$ in eq.(\ref{ints}) are
identified with the representations of the operators $\CO=\CO(A,P)$
in the $A,B$ variables:
\debut
L_\CO(A_1,\cdots,A_n|B_1,\cdots,B_n) \Longleftrightarrow \CO\({A,P(A,B)}\) 
\non
\fin
This is particularly clear for the primary operators
$\Phi_m= e^{im\varphi}$. 
This observation actually underlyies the construction we 
describe in the following.

\section{KdV equation and hyperelliptic curves.}
The ultraviolet limit of the (restricted) Sine-Gordon model
is a minimal conformal field theory. Its classical
limit is therefore intimitely related to the KdV equation.
One may think of KdV as describing one of the chiral
sector of Sine-Gordon.
In this section, we first present a new description of
the space of local fields in KdV in terms of the local
integrals of motion and their densities. We then describe
various connexion between form factor formula
and hyperelliptic curves and the associated
finite zone solutions of KdV.

\subsection{Local fields and null vectors in the KdV theory.}
The KdV equation for a field $u(t_1, t_3, \cdots)$
is the following non linear equation:
\begin{eqnarray}
\partial_{3} u +{3\over 2}uu' -{1\over 4} u''' =0 \label{kdveq}
\end{eqnarray}
We shall use both notations $\partial _1$ and $'$ for the
derivatives with respect to $x=t_1$.
As is well known this is one of a hierarchy of
equations which can be written in a Lax form. Namely the
field $u$ depends on a set of time variables $t_{2k-1}$,
and its evolution with respect to these times is
encoded in the equations~: 
\begin{eqnarray}
{\partial L \over \partial t_{2k-1}}
&=& \left[ \left( L^{2k-1\over 2} \right)_+, L
\right] = {1 \over 2^{2k-1}}u^{(2k-1)}+ \cdots   \label{motionkdv}
\end{eqnarray}
Here $L$ is the Lax operator of KdV~:
\debut
L = \partial_1^2 - u \label{laxkdv}
\fin
We have used the pseudo-differential
operator formalism of Gelfand and Dickey, cf. \cite{dickey}.
 
In the KdV theory, the local fields, which are the descendents 
of the identity operators, are
simply polynomials in $u(t)$ and its derivatives  with respect to $t_1$:
\begin{eqnarray}
\CO &=& \CO (u,u',u'',...)        \label{loc1}
\end{eqnarray}
Instead of the variables $u,u',u'',...$, we may replace
the odd derivatives of $u(x)$ by the higher time derivatives $\partial_{2k-1} u$,
according to the equations of motion of the hierarchy (\ref{motionkdv}).
We may also replace the even derivatives of $u(x)$ by the densities 
$S_{2k}$ of the local integrals of motion,
\begin{eqnarray}
S_{2k} &=& res_{\partial _1} L^{2k-1\over 2} =-{1\over 2^{2k-1}} u^{(2k-2)}+
\cdots,  \nonumber \end{eqnarray}
In particular $S_2=-{1\over 2}u$.
For a reader who prefers the $\tau$-function language $S_{2k}=\partial_1
\partial _{2k-1} \log \tau$. They satisfy the conservation laws~:
$\d_{2l+1}S_{2k}=\d_1H_{2k+2l}$ for some local field $H_{2k+2l}$.
Therefore, from analogy with the conformal case we suggested in \cite{bbs2}
the following conjecture~:
 
\proclaim Conjecture.
We can write any local fields of the KdV theory as
\begin{eqnarray}
{\cal O} (u,u',u'',...) &=& F_{{\cal O},0} (S_2, S_4, \cdots) +
\sum_{\nu \geq 1}\partial^\nu F_{{\cal O},\nu} (S_2, S_4, \cdots)
\label{loc2}
\end{eqnarray}
where $\nu =( i_1,i_3, \cdots)$ is a multi index,
$\partial^\nu = \partial^{i_1}_{1} \partial^{i_3}_{3} \cdots $,
$|\nu | = i_1 + 3 i_3 +\cdots$.
 
We checked this conjecture up to very high levels.
To see that this conjecture is a non trivial one, 
let us compute the character $\chi_1$
of the space of local fields eq.(\ref{loc1}). Attributing the degree 2 to
$u$ and 1 to $\partial_1$, we find that~:
\begin{eqnarray}
\chi_1 &=& \prod_{j \geq 2} {1 \over 1 - p^j} =
(1-p) \prod_{j \geq 1} {1 \over 1 - p^j}
= 1 +p^2 +p^3 +2p^4 +2p^5 +\cdots
\nonumber
\end{eqnarray}
On the other hand the character $\chi_2$ of the elements 
in the right hand side of eq.(\ref{loc2}) is~:
\begin{eqnarray}
\chi_2 &=& \prod_{j\geq 1} {1 \over 1-p^{2j-1}}\prod_{j\geq 1} {1 \over 1-p^{2j}}
= \prod_{j \geq 1} {1 \over 1 - p^j}
= 1 +p +2p^2 +3p^3 +5 p^4 +7p^5 +\cdots
\nonumber
\end{eqnarray}
Hence $\chi_1 < \chi_2$ and the two spaces in eq.(\ref{loc2})
can be equal only if there are null-vectors among the elements
in the right hand side of eq.(\ref{loc2}).
Let us give some examples of null-vectors~:
\begin{eqnarray}
level~1&:& \partial_{1}\cdot 1 =0 \label{nvcl} \\
level~2&:& \partial^2_{1}\cdot 1 =0 \nonumber \\
level~3&:& \partial^3_{1}\cdot 1 =0, \quad\partial_{3}\cdot 1 =0 \nonumber \\
level~4&:& \partial^4_{1}\cdot 1 =0,
\quad \partial_{1}\partial_{3}\cdot 1 =0,
%\nonumber \\&&
\quad(\partial^2_{1} S_2 -4 S_4 +6 S_2^2)\cdot 1 =0 \nonumber \\
level~5&:& \partial^5_{1}\cdot 1 =0,\quad\partial_{1}^2\partial_{3}\cdot 1 =0,
\quad\partial_{5}\cdot 1 =0, \nonumber \\
&&\partial _{1}(\partial^2_{1} S_2 -4 S_4 +6 S_2^2)\cdot 1 =0,
\quad (\partial_{3} S_2 -\partial_{1} S_4 )\cdot 1 =0 \non
\end{eqnarray}
We wrote all the null-vectors
explicitly to show that their numbers exactly
match the character formulae.
The non trivial null-vector at level 4 expresses $S_4$
in terms of the original variable $u$:
$4 S_4=-{1\over 2}u'' +{3\over 2} u^2 $. With this identification
the non-trivial null-vector at level 5, $\partial_{3} S_2 -\partial_{1} S_4$,
gives the KdV equation itself.

In summary, null vectors code the hierarchy of equations of motion.

\subsection{Hyperelliptic curves and Riemann bilinear identity.}
Let us consider an hyperelliptic curve
$\Gamma$ of genus $n$ described by the equation
\debut
\Gamma~: ~~~Y^2 &=& X {\cal P}(X),~~\quad
with\quad {\cal P}(X) = \prod_{j=1}^{2n}(X -B_j^2) ,
\label{hyper}
\fin
We suppose that the coefficients $B_i$ have been ordered~:
$ B_{2n}>\cdots >B_2>B_1>0$.
For historical reasons we prefer to work with the parameter $A$
such that $X=A^2$. Thus the curve $\Ga$ is~:
\debut
\Ga~:~~~Y^2 = A^2 ~\CP(A^2) \non
\fin 
The surface is realized as the $A$-plane with cuts
on the real axis over the intervals
$c_i=(B_{2i-1}, B_{2i})$ and $\bar{c}_i=(-B_{2i}, -B_{2i-1})$,
$i=1,\cdots, n$,
the upper (lower) bank of $c_i$ is identified with
the upper (lower) bank of $\bar{c}_i$.
The  square root $\sqrt{\CP(A^2)}$ is chosen so that
$\sqrt{\CP(A^2)}\to A^{2n}$ as $A\to\infty$.
The canonical basis of cycles is chosen as follows:
the cycle $a_i$ starts from $B_{2i-1}$ and goes in the upper half-plane
to $-B_{2i-1}$, while the cycle $b_i$ is an 
anti-clockwise cycle around the cut $c_i$.
 
Since $\Ga$ has genus $n$, there are $n$ independent holomorphic
differentials on it. A basis is given by
$d\sigma_k(A) = {A^{2k-2} \over \sqrt{{\cal P}(A^2)}}dA$,
for $k=1,\cdots, n $.
The normalized holomorphic differentials $d\omega _i$ for
$i=1,\cdots ,n$ are such that~:
\debut
\int\limits _{a_j}d\omega _i =\delta _{i,j}, \non
\fin
They are linear combinations of the $d\sigma_k(A)$ with coefficients
depending on $B_i$.  They can written as $n\times n$ determinants~:
\debut
d\om_k(A) = \inv{\De} \det ~M(A)\quad with\quad
\cases{ M(A)_{ij}= 
\int\limits_{a_i} \frac{D^{2(j-1)}}{\sqrt{\CP(D^2)}}dD, &if $i\not= k$,\cr
M(A)_{kj} = {A^{2j-2} \over \sqrt{{\cal P}(A^2)}}dA ,&
$i,j=1,\cdots,n$.\cr}
\non
\fin
with 
\debut
\De = \det\({ 
\int\limits_{a_i} \frac{D^{2(j-1)}}{\sqrt{\CP(D^2)}}dD}\)_{i,j=1,\cdots,n} 
\label{normdelta}
\fin

A particular role is also played by the differentials of the second
kind with singularities at infinity. These are meromorphic differentials
whose only singularities are poles of order bigger or equal to two
at infinity. Such differentials are linear combinations of
differentials of the form $\frac{A^{2n+2p}}{\sqrt{\CP(A^2)}} dA$ with 
$p\geq 0$.  The normalized second kind differentials 
with singularity at infinity $d\tilde{\omega} _{2i-1} $, $i\ge 1$
are defined by~:
\debut
\int\limits _{a_j} d\tilde{\omega} _{2i-1}=0,\quad and \quad
d\tilde{\omega} _{2i-1}(A)=d(A^{2i-1})+\CO(A^{-2})dA ~~~ for ~~~ A\sim\infty
\non%\label{diffs}
\fin

%An explicit expression is~:
%\debut
%d\tilde{\omega} _{2p+1}(A) =(2p+1) \Delta ^{-1}\ det (M(A)) dA\non
%\fin 
%with $M(A)$ is $(n+1)\times (n+1)$ matrix with the following matrix
%elements
%\debut
%&&M(A)_{i,j}=\int\limits _{a_j}{D^{2(i-1)}\over \sqrt{\CP(D^2)}} dD,
%\quad i,j=1,\cdots ,n \non\\
%&&M(A)_{i,n+1}={A^{2(i-1)}\over \sqrt{\CP(A^2)}}, \quad i=1,\cdots ,n \non\\
%&&M(A)_{n+1,j}= \int\limits _{a_j}{Q_p(D^2)\over \sqrt{\CP(D^2)}},
%\quad j=1,\cdots ,n \non\\
%&&M(A)_{n+1,n+1}= {Q_p(A^2)\over \sqrt{\CP(A^2)}}
%\non
%\fin
%where
%$$
%Q_p(A^2)=\int\limits _{|D|>|A|}
%dD {D^{2p+1}\over D^2-A^2} \sqrt{\CP(D^2)}=
%\[ \sqrt{\CP(A^2)}A^{2p}\]_+
%$$
%where $[\cdots]_+$ means taking the polynomial part in the
%expansion around infinity.

On Riemann surfaces there is a natural symplectic pairing
between meromorphic differentials. Namely, 
let $d\Om_1$ and $d\Om_2$ be two meromorphic differentials on $\Ga$.
The pairing $(d\Om_1\bullet d\Om_2)$ is then defined by integrating 
them along the canonical cycles as follows~:
\debut
(d\Om_1 \bullet d\Om_2) = \sum_{i=1}^n\( 
\int\limits _{a_j} d\Om_1 \int\limits _{b_j} d\Om_2 
-\int\limits _{a_j} d\Om_2 \int\limits _{b_j} d\Om_1 \)
\non
\fin
The Riemann bilinear identity expresses this quantity
in terms of sum over residues~:
\debut
(d\Om_1 \bullet d\Om_2) = \inv{2i\pi} \sum_{poles}
res( \Om_1 d\Om_2) \label{riemclass}
\fin
In particular, the pairing between the 
normalized holomorphic differentials
is trivial~: $(d\om_i\bullet d\om_j)=0$ for $i,j=1,\cdots,n$.

As formulated in the previous equations, the
Riemann bilinear identity gives an expression
for the pairing between one-forms. We now
want to formulate it in a dual form, ie. in
a form which gives an expression for the pairing between
one-cycles. More precisely, let $C_1$ and $C_2$ be
two cycles, the pairing is simply the intersection number~:
\debut
(C_1\circ C_2) = 
\sum_{j=1}^n\( n_j^1 m_j^2 - m_j^1 n_j^2 \) \label{definter}
\fin
if $C_1= \sum_{j=1}^n\(n_j^1 a_j + m_j^1b_j\)$, and similarly for $C_2$.
The dual form of the bilinear Riemann identity is:

\proclaim Proposition.
Let $d\om_j$ be the normalized holomorphic differentials.
Let $d\xi_j$, for $j=1,\cdots,n$, be differentials of the second 
kind dual to the holomorphic differentials, ie. such that
$$ (d\om_i\bullet d\xi_j)=\de_{ij},\quad and\quad
(d\xi_i\bullet d\xi_j) =(d\om_i\bullet d\om_j) =0 $$
Then the intersection number between two cycles $C_1$ and $C_2$
can be written as~:
\debut
(C_1\circ C_2)= \sum_{j=1}^n\( \int\limits_{C_1} d\om_j
\int\limits_{C_2}d\xi_j - \int\limits_{C_2} d\om_j
\int\limits_{C_1}d\xi_j \)
\label{riemdual}
\fin
Alternatively, the intersection number is given by~:
\debut
(C_1\circ C_2)= \inv{2i\pi}
\int\limits _{C_1}{dA_1\over\sqrt{\CP (A_1^2)}}
\int\limits _{C_2}{dA_2\over\sqrt{\CP (A_2^2)}}\ C_{cl}(A_1,A_2)
\label{Cclass}
\fin
where the anti-symmetric polynomial $C_{cl}(A_1,A_2)$ is given by
\debut
C_{cl}(A_1,A_2)=
\sqrt{\CP (A_1^2)}{d\over dA_1}\( \sqrt{\CP (A_1^2)} {A_1\over A_1^2-A_2^2}\)
- (A_1 \longleftrightarrow A_2) \label{defCclass}
\fin
\par
\proof
See, for example, \cite{sm3} for a relevant discussion.
The normalization condition for the differentials $d\om_j$
and $d\xi_j$ means that the matrix $P$ defined by,
\debut
P_{ij} = \pmatrix{ \int\limits_{a_j} d\om_i & \int\limits_{b_j} d\om_i \cr
	\int\limits_{a_j} d\xi_i & \int\limits_{b_j} d\xi_i \cr} \non
\fin
is a symplectic matrix. ie:
\debut
P~J~ {}^t P = J\quad with \quad J=\pmatrix{0 &  id \cr -id & 0\cr}
\label{relpourP}
\fin
where ${}^t P$ denotes the transposed matrix. 
Notice that since $J^2=-id$, eq.(\ref{relpourP}) means that the 
right inverse of $P$ is $-J~ {}^t P~J$. Using the fact the right
and left inverse are identical, eq.(\ref{relpourP}) is therefore
equivalent to ${}^t P~J~P= J$.

Now let $C_1$ and $C_2$ be our two cycles. By definition
the intersection number is $(C_1\circ C_2) = \bra{C_1}J\ket{C_2}$,
where $\bra{C_1}=(n_j^1,m_j^1)$ and similarly for $\ket{C_2}$.
Using the relation ${}^t P~J~P= J$, we can rewrite the
intersection number as~:
\debut
(C_1\circ C_2)= \bra{C_1} {}^t P~J~P\ket{C_2} \non
\fin
This is equivalent to the relation (\ref{riemdual}) since the vector 
$\bra{C_1} {}^t P $ is the vector of the periods of the forms $d\om_j$ and
$d\xi_j$ along the cycles $C_1$~: 
\debut
\bra{C_1} {}^t P = (\int\limits_{C_1}d\om_j, \int\limits_{C_1}d\xi_j)\non
\fin
and similarly for $P\ket{C_2}$.

The second formulation (\ref{Cclass}) can be proved in two ways.
Either one verifies directly that the integral (\ref{Cclass})
gives the intersection numbers (the integral is localized
on the intersection of the cycles), and then by expanding $C_{cl}(A_1,A_2)$
this gives a formula for the differentials $d\xi_j$. Indeed,
the explicit expression of $C_{cl}(A_1,A_2)$ is~:
\debut
C_{cl}(A_1,A_2)=\inv{A_1^2-A_2^2}\({ A_1^2\CP'(A_1^2) + A_2^2\CP'(A_2^2)
-\frac{A_1^2+A_2^2}{A_1^2-A_2^2}(\CP(A_1^2)-\CP(A_2^2)) }\) \non
\fin
It is an anti-symmetric polynomials of degree at most $4n-2$. It 
can be expanded as:
\debut
C_{cl}(A_1,A_2)= \sum_{k=1}^n\( A^{2k-2}_1 Q_k(A_2^2) - 
A^{2k-2}_2 Q_k(A_1^2) \) \non
\fin
where $Q_k(A^2)$ are polynomials of degree $(4n-2k)$ given by, 
\debut
Q_k(A^2)=\sum_{p= 2k}^{2n} (-)^{p+1} (p+1-2k)s_{4n-2p}(B) A^{2p-2k} \non
\fin
These polynomials define the differentials of the second kind 
dual to the holomorphic forms $d\sig_k(A)$. Alternatively,
one may determine directly the differentials $d\xi_j$ by solving
their normalization conditions, and then by resumming
$\sum_j d\om_j\wedge d\xi_j$ this gives the formula for $C_{cl}(A_1,A_2)$.
\square

It is the dual formulation of the Riemann bilinear identities
which admits a simple quantum deformation in the form factor problem
\cite{sm3}.

\subsection{Baker-Akhiezer functions and finite zone solutions.}
As is well known, to any hyperelliptic curve we can associate a solution
of the KdV equation. 
We first need certain informations about the Baker-Akhiezer function.
The Baker-Akhiezer function $w(t,A)$ is a eigenfunction
of the Shroedinger equation defined by $L$
with eigenvalue $A^2$, 
\begin{eqnarray}
L ~~w(t,A) = A^2~~ w(t,A)
\label{baker}
\end{eqnarray}
which admits an asymptotic expansion at $A = \infty$ of the form
\begin{eqnarray}
w(t,A) = e^{\zeta(t,A)}\({ 1 + 0(\inv{A})}\);~~~\quad with\quad \zeta(t,A) 
= \sum_{k\geq 1} t_{2k-1} A^{2k-1}
\nonumber
\end{eqnarray}
In these formulae, higher times are considered as parameters.
The second solution of equation (\ref{baker}), denoted
by $w^*(t,A)$, has the asymptotics
\begin{eqnarray}
w^*(t,A) = e^{-\zeta(t,A)}~\({ 1 + 0(\inv{A})}\)
\nonumber
\end{eqnarray}
These definitions do not fix completely the Baker-Akhiezer
functions since we can still multiply them by constant
asymptotic series of the form $(1+O(1/A))$. Since
normalizations will be important,
let us give a more precise definition.
We first introduce the dressing operator $\Phi$, which is an
element of the algebra of pseudo-differential operators, by~:
\debut
L = \Phi \partial_1^2 \Phi^{-1};
\quad with\quad \Phi = 1 +\sum_{i>1} \Phi_i \partial_1^{-i}
\non
\fin
We then define the Baker-Akhiezer functions by~:
\debut
w(t,A) = \Phi e^{\zeta(t,A)},\quad and\quad
w^*(t,A) = (\Phi^*)^{-1} e^{-\zeta(t,A)}
\non
\fin
where $\Phi^* = 1 +\sum_{i>1} (-\partial_1)^{-i} \Phi_i$ is the formal adjoint of
$\Phi$. Clearly, $w(t,A)$ is a solution of the Shroedinger equation
with eigenvalue $A^2$. Moreover \cite{bbs2}, 
\proclaim Proposition.
With the above definitions, one has \hfill \break
\noindent
1) The wronskian $W(A)=w(t,A)' w^*(t,A) - w^*(t,A)'w(t,A)$ takes the value
$W(A) = 2 A$.\hfill\break
\noindent
2) The generating function of the local densities $S(A)=1+\sum_{k>0}
S_{2k}A^{-2k}$ is related to the Baker-Akhiezer function by
\debut
S(A) = w(t,A) w^*(t,A)        \label{Swwstar}
\fin
%3) The function $S(A)$ satisfies the Ricatti equation
%\debut
%2S(A)S(A)''-(S(A)')^2-4uS(A)^2-4A^2S(A)^2+4A^2=0
%\label{ric} \fin
\par

The solutions of KdV associated to hyperelliptic curves are the 
so-called finite-zone solutions. The Baker-Akhiezer function is then 
an analytical function on the spectral curve.
Let us recall briefly the construction \cite{itsmat,novbook}.
Consider the hyperelliptic curve (\ref{hyper}) which we have
introduced in the previous section. 
Let us consider in addition a divisor of order $n$ on the surface $\Gamma$:
$$\CD =(P_1,\cdots ,P_n)$$
With these data we construct the Baker-Akhiezer function which is the unique
function with the following analytical properties:
\begin{itemize}
\item It has an essential singularity at infinity:
$w(t,A) = e^{\zeta(t,A)}(1 + O(1/A))$.
\item It has $n$ simple poles outside infinity. The divisor of these
poles is $\CD$.
\end{itemize}
Considering the quantity $-\partial_1^2 w + A^2 w$, we see that
it has the same analytical properties as $w$ itself, apart for the first
normalization condition. Hence, because $w$ is unique,
 there exists a function $u(t)$ such that
\debut
-\partial_1^2w + u(t) w + A^2 w =0
\label{A0}
\fin
We recognize eq.(\ref{baker}).
One can give various explicit constructions of the Baker-Akhiezer function.
Let us introduce the divisor  ${\cal Z}(t)$
of the zeroes of the Baker-Akhiezer
function. It is of degree $n$:
\debut
\CZ(t)  = (A_1 (t), \cdots, A_n(t) ) \non
\fin
The equations of motion with respect to the first time
for the divisor $\CZ(t)$ read \cite{novbook}:
\debut
\partial_1
A_i(t) = -{  \sqrt{ {\cal P}(A_i^2(t))} \over \prod\limits_{j \neq i}
(A_i^2(t) - A_j^2(t) ) }
\label{A5}
\fin
The normalization of the Baker-Akhiezer function corresponds to
a particular choice of the divisor of its poles $\CD$.
We shall specify the divisor which corresponds to the
normalization of the Baker-Akhiezer function which
was required above.
We have the following proposition \cite{bbs2}.
 
\proclaim Proposition.
For the Baker-Akhiezer functions $w(t,A)$ and $ w^*(t,A)$ normalized
such that their Wronskian is $2A$, ie. 
$w(t,A)' w^*(t,A) - w^*(t,A)' w(t,A)=2A $.  We have~:
\begin{eqnarray}
S(A) = {Q(A^2) \over \sqrt{{\cal P}(A^2)} } \label{Shyper}
\fin
where the polynomials $Q(A^2)$ and ${\cal P}(A^2)$ are~: 
$Q(A^2) =\prod\limits_{i=1}^n (A^2 - A_i^2)$ and 
${\cal P}(A^2) = \prod\limits_{i=1}^{2n} (A^2 - B_i^2)$. 
\par

We now can make contact with the generating function (\ref{gf1})
for the form factors of the descendent operators. Indeed, let
us introduce a set of variables $J_{2k}$ related to the generating
function $S(A)$ by~:
\debut
S(A) \equiv
 \exp \left(- \sum_k {1\over k} J_{2k}A^{-2k} \right)\label{deffJ}
\fin
The formula (\ref{Shyper}) gives~:
\begin{eqnarray}
J_{2k} = \sum_i A_i^{2k} -{1\over 2} \sum_i B_i^{2k}
\nonumber
\end{eqnarray}
They coincide with those appearing when defining form factors
of descendents operators, cf. eq.(\ref{defJ}).
Clearly the quantities $I_{2k}=\sum_i B_i^{2k}$ coincide with the
value of the integral of motion for the finite zone solutions.
In other words, the generating function of the descendents operators
(\ref{gf1}) is in correspondence with the integrals of motion 
and their densities for the finite zone solutions. Moreover,
the variables $A,B$ used in the integral representation of the
form factors are in correspondence with the poles and zeroes
of the Baker-Akhiezer functions.

\subsection{The ultra-classical limit of the form factor formula.}

There is a surprising relation between the form factor formula
and the averaging formula occuring in 
the Whitham theory for KdV \cite{whi,fla,kri,dn}. 
The present section is devoted to the description of this relation.

Let us remind briefly what is the Whitham method about.
Suppose we consider the solutions of
KdV which are close to a given quasi-periodic solution.
The latter is defined by the set of ends of zones
$B_1^2,\cdots ,B_{2n}^2$. We know that for the finite-zone
solution the dynamics is linearized by the Abel transformation
to the Jacobi variety of the hyper-elliptic surface
$Y^2=X\CP (X)$ for $\CP (X)=\prod (X-B_j^2)$.
The idea of the Whitham method is to average over the fast
motion over the Jacobi variety and to introduce "slow times"
$T_j$ which are related to the original KdV times
as $ T_j=\epsilon t_j$ ($\epsilon \ll 1$), assuming that the
ends of zones $B_j$ become functions of these "slow times"
(recall that the ends of zones were the integrals of motion for the pure
finite-zone solutions).

For the given finite-zone solution the observables can be written in
terms of $\theta$-functions on the Jacobi variety, but this kind
of formulae is inefficient for writing the averages. One has
to undo the Abel transformation, and to write the observables in
terms of the divisor $\CZ=(A_1,\cdots ,A_n)$. The formulae for the
observables are much more simple in these variables, and the averages
can be written as abelian integrals, the Jacobian due to the Abel transformation
is easy to calculate. The result of this calculation is as follows \cite{fla}.
Every observable $\CO$ can be written as an even symmetric function
$L_{\CO}(A_1,\cdots ,A_n)$
(depending on $B$'s as parameters). For the average we have
\debut
\vev{\langle \ \CO\ \rangle}=\Delta ^{-1}
\int\limits _{a_1}{dA_1\over\sqrt{\CP(A^2_1)}}\cdots
\int\limits _{a_n}{dA_n\over\sqrt{\CP(A^2_n)}} L_{\CO}(A_1,\cdots ,A_n)
\prod\limits _{i<j}(A_i^2-A_j^2)
\label{whithamdef}
\fin
where the normalization factor $\De$ is defined as above in eq.(\ref{normdelta}).
%$$\Delta=det \(\int\limits _{a_i}{A^{2j-2}dA\over\sqrt{\CP(A^2)}}\)_
%{i,j=1,\cdots n}$$

The similarity of this formula with the formula for the form factors
(\ref{ff}) is a surprising fact.  We have the following dictionary: \\
\noindent For the local observables, we have
\debut
L_{\CO} \Longleftrightarrow L_{\CO}
\non
\fin
For the weight of integration, we have
\debut
{1\over \sqrt{{\cal P}(A^2)}} \Longleftrightarrow \prod_{j=1}^{2n} \psi(A,B_j)
\non
\fin
For the integration cycles, we have
\debut
a_i-cycles \Longleftrightarrow functions~~ A_i^{2\nu}=a_i \non
\fin
The most striking feature is that the cycles of integration are replaced
by functions of $a_i = A_i^{2\nu}$.
The coincidence between the notations for $a_i$-variables and
$a_i$-cycles is therefore not fortuitous.
The explanation of the fact that the cycles are replaced by these functions is
given in \cite{bbs}, where it was shown that the factor
$\prod_i a_i^{-i}$ selects the
classical trajectory in the semi-classical approximation of eq.(\ref{ints}).
So, the solution of a non trivial, full fledged, quantum field theory has
provided us with a very subtle definition of a {\it quantum} Riemann surface.

For comparison with the quantum case, it is important to note
that the average (\ref{whithamdef}) can vanish for some
observables $L_\CO(A_1,\cdots,A_n)$. More precisely, let
$M_\CO(A_1,\cdots,A_n)$ be the antisymmetric polynomials defined by~:
\debut
M_\CO(A_1,\cdots,A_n)= \prod_{i<j}(A^2_i-A_j^2)~L_\CO(A_1,\cdots,A_n) 
\non
\fin
The degree of $M_\CO$ in any variable $A_k$ is always greater than $(2n-2)$.
Since $M_\CO(A_1,\cdots,A_n)$ are even antisymmetric polynomials, a basis
of such functions is provided by $n\times n$ determinants of $n$ polynomials
$Q_{p_j}(A^2_k)$ of degree $(2p_j-2)$. The average $\vev{\vev{\CO_{\{p_j\}}}}$
is then~:
\debut
\vev{\vev{\CO_{\{p_j\}}}}= \De^{-1} ~~\det\({ \int\limits_{a_i}
\frac{dA}{\sqrt{\CP(A^2)}} Q_{p_j}(A^2) }\)_{i,j=1,\cdots,n} \non
\fin
For certain observable $\CO$ the average $\vev{\vev{\CO}}$ vanishes.
There are two origins for this vanishing~:

{\bf 1. Exact forms.} The integral (\ref{whithamdef}) vanishes if
$M_\CO(A_1,\cdots,A_n)$ is an exact form, ie.:
\debut
M_\CO(A_1,\cdots,A_n)= \sum_k(-)^k \hat M(A_1,\cdots,\hat A_k,\cdots,A_n)
\times Q(A^2_k)\non
\fin
such that the differential $\frac{Q(A^2)}{\sqrt{\CP(A^2)}}dA$ has 
vanishing integrals along the $a_i$-cycles, 
and for some antisymmetric polynomial $M(A_1,\cdots,A_{n-1})$.
Here $\hat A_k$ means that the variables $A_k$ is omitted.

{\bf 2. Riemann bilinear identity}. Since we are integrating on $n$ 
non-intersecting cycles, the integral (\ref{whithamdef}) vanishes if:
\debut
M_\CO(A_1,\cdots,A_n)=\sum_{i<j} \hat M(A_1,\cdots,\hat A_i,\cdots,\hat A_j,
\cdots,A_n)~ C_{cl}(A_i,A_j) \non
\fin
where $C_{cl}(A_1,A_2)$ is defined in (\ref{defCclass}), and
$\hat M(A_1,\cdots,A_{n-2})$ is an anti-symmetric polynomial. This fact is a direct
consequence of the dual form of the Riemann bilinear identities.

The null-vectors of the quantum theory originate in the
quantum deformation of these two properties.
 
\section{The deformed Riemann bilinear identity and null-vectors.}
We now describe how a quantum deformation of the geometrical structure 
we just recalled leads to the notion of null vectors in integrable
field theory. The existence of these null vectors yields differential
equations for the correlation functions or for the form factors, which
reduce to the KdV hierarchy in the classical limit.

%\subsection{Deformed Riemann bilinear identity.}
\subsection{Null vectors in integrable field theory.}
By definition, null-vectors correspond to operators with all the form factors
vanishing. Therefore, consider the fundamental integrals $\hat f_\CO$
of the form factor formula~:
\debut
{1\over (2\pi i)^n}\int dA_1\cdots \int  dA_n
\prod\limits _{i=1}^n \prod\limits _{j=1} ^{2n} \psi (A_i,B_j)
\prod\limits _{i<j} (A_i^2-A_j^2)
\ L_\CO^{(n)} (A_1,\cdots ,A_n|B_1,\cdots ,B_{2n})
\prod\limits _{i=1}^n a_i^{-i} \label{inte}
\fin
Instead of $L_\CO^{(n)}$, it is more convenient to use the anti-symmetric
polynomials $M_\CO^{(n)} $:
\debut
&&M_\CO^{(n)} (A_1,\cdots ,A_n|B_1,\cdots ,B_{2n})=
%\non\\&&\hskip 0.5cm =
\prod\limits _{i<j} (A_i^2-A_j^2)
\ L_\CO^{(n)} (A_1,\cdots ,A_n|B_1,\cdots ,B_{2n})
\non \fin
The dependence
on $B_1,\cdots ,B_{2n}$ in the polynomials  $M_\CO^{(n)} $ will
often be omitted.
 
There are several reasons why this integral can vanish. Some of
them depend on a particular value of the coupling constant or
on a particular number of solitons. We should not consider these
occasional situations. In parallel to the classical case discussed above,
there are three general reasons for
the vanishing of the integral, let us present them.
 
{\bf 1. Residue.} The integral (\ref{inte}) vanishes if vanishes the residue
with respect to $A_n$ at the point $A_n=\infty$
of the expression
$$
\prod\limits _{j=1} ^{2n} \psi (A_n,B_j)a_n^{-n}
M_\CO^{(n)} (A_1,\cdots ,A_n)
$$
Of course the distinction of the variable $A_n$ is of no importance
because $M_\CO^{(n)} (A_1,\cdots ,A_n)$ is anti-symmetric.
 
{\bf 2. "Exact forms."}
The integral (\ref{inte}) vanishes if
$M_\CO^{(n)} (A_1,\cdots ,A_n)$ happens to be
an "exact form". Namely, if it can be written as:
\debut
&&M_\CO^{(n)} (A_1,\cdots ,A_n) =\sum\limits _k (-1)^k
M(A_1,\cdots ,\widehat{A_k},\cdots ,A_n)
\(Q(A_k)P(A_k)-qQ(qA_k)P(-A_k)\),
\label{zero}
\fin
with
$$ P(A)=\prod_{j=1}^{2n} (B_j+A) $$
for some anti-symmetric polynomial $M(A_1,\cdots ,A_{n-1})$.
Here $\widehat{A_k}$ means that $A_k$ is omitted.
The polynomial $P(A)$ should not be confused with the
polynomial $\CP(A^2)$. They are related by:
$\CP(A^2)= P(A)P(-A)$.
Eq.(\ref{zero}) is a direct consequence of the functional equation satisfied by
$\psi(A,B)$~:
\debut
\psi(qA,B)= \({\frac{B-A}{B+qA} }\) \psi(A,B) \label{eqpsi}
\fin 
For $Q(A)$ one can take in principle any Laurent polynomial,
but since we want $M_\CO^{(n)}$ to be a polynomial
the degree of $Q(A)$ has to be greater or equal $-1$.

{\bf 3. Deformed Riemann bilinear relation.} The integral
(\ref{inte}) vanishes if
\debut
M_\CO^{(n)} (A_1,\cdots ,A_n)=
\sum\limits _{i<j}(-1)^{i+j}M(A_1,\cdots ,\widehat{A_i},
\cdots ,\widehat{A_j},\cdots A_n)C(A_i,A_j)
\non \fin
where $M(A_1,\cdots ,A_{n-2})$ is an anti-symmetric polynomial
of $n-2$ variables, and $C(A_1,A_2)$ is given by
\debut
C(A_1,A_2 )={1\over A_1A_2}\left\{ {A_1-A_2\over A_1+A_2 }
(P(A_1)P(A_2)-P(-A_1)P(-A_2))
+
(P(-A_1)P(A_2)-P(A_1)P(-A_2))\right\}  \label{C}
\fin
This property needs some comments. For the case of generic coupling constant
its proof is rather complicated. It is a consequence of the so
called deformed Riemann bilinear
identity \cite{sm2}.
The name is due to the fact that
in the limit $\xi\to\infty$ (which is the opposite of the
classical limit which corresponds to $\xi\to 0$) the deformed
Riemann bilinear identity happens to be
the same as the Riemann bilinear identity for hyper-elliptic integrals.
The formula for $C(A_1,A_2)$ given in \cite{sm3} differs from (\ref{C})
by simple "exact forms". Notice that the formula for $C(A_1,A_2)$
does not depend on the coupling constant.
For the reflectionless case a
very simple proof is available.

\proclaim Proposition.
The function $C(A_1,A_2)$ defined in eq.(\ref{C}) satisfy~:
\debut
\int dA_1 \int dA_2
\prod\limits _{i=1}^2 \prod\limits _{j=1} ^{2n} \psi (A_i,B_j)
C(A_1,A_2 ) a_1^{k}a_2^{l}=0\quad  ~~~~~~\forall k,l \label{ii}
\fin
\par
 
\proof
The reflectionless case is a rather degenerate one, so, the
deformed Riemann bilinear identity \cite{sm2} does not exist
in complete form. However we only use
the consequence of the deformed Riemann bilinear identity which
allows a simple proof in the reflectionless case. 
Let us introduce the functions
$$F(A)=\prod\limits _{j=1} ^{2n} \psi (A,B_j)P(A),
\qquad G(A)= \prod\limits _{j=1} ^{2n} \psi (A,B_j)P(-A) $$
Recall that the function $\psi(A,B)$ satisfies the difference
equation (\ref{eqpsi})~:
$\psi(Aq,B)=\left({B-A\over B+qA}\right)~ \psi(A,B)$.
It implies that
$$F(Aq)=G(A) $$
The integral (\ref{ii}) can therefore be rewritten as follows:
\debut
\int {dA_1\over A_1} \int {dA_2 \over A_2}
%&&{1\over A_1A_2}
%\non\\&&\times
\left\{ {A_1-A_2\over A_1+A_2 }
(F(A_1)F(A_2)-F(qA_1)F(qA_2))+
(F(qA_1)F(A_2)-F(A_1)F(qA_2))\right\}
a_1^{k}a_2^{l}  \non
\fin
Changing variables $A_i\to qA_i$ where needed one easily find
that this integral equals zero. Recall that $a_i=A_i^{2\nu}$ and
$q^{2\nu}=1$, so $ a_1^{k}a_2^{l} $ do not change under these changes of
variables.
\square

\subsection{The deformed Riemann bilinear identity.}
As understood in \cite{sm2}, the complete structure
underlying the deformed bilinear identity only emerges when one
consider the general case. Ie. one has to consider the
Sine-Gordon model at generic coupling constant $\xi=\pi/\nu$.
There is then two dual quantum parameters $q$ and $\tau$:
\debut
q = e^{i\frac{\pi}{\nu}}\quad;\quad \tau= e^{i\pi\nu} \non
\fin
The basic ingredient in the form factors at generic value of
the coupling constant is a (special) pairing 
between polynomials $L(A)$ and $r(a)$ with $a=e^\al$
and $a=A^{2\nu}=e^{2\nu\al}$ for some $\al$.
The pairing is defined as:
\debut
\vev{L(A), r(a)} = (\int\limits_{quant.}\cdots)~ L(A)r(a)
\label{paires}
\fin
where the package $(\int\limits_{quant.}\cdots)$ refers to
some very complicated contour integrals \cite{book} which
reduce to those involved in eq.(\ref{ints}) in the reflexionless
cases.

The main lesson from ref.\cite{sm2} is that this pairing could
be understood as the analogue of the pairing between one-forms
and one-cycles simply defined by integrating the one-form, say $d\om$,
along the one-cycle, say $C$~:
\debut
\vev{L(A), r(a)} \Longleftrightarrow \int\limits_C ~d\om \non
\fin
Under this analogy one has the following
possible identification:
\debut
forms &\Longleftrightarrow& L(A),
\qquad ~~~~around\quad\nu\simeq\infty \non\\
cycles &\Longleftrightarrow& r(a) \non
\fin
This identification is appropriate close to the semi-classical
limit $\nu\to\infty$, as we hope to have convinced the reader.
But in the opposite limit, this is the dual identification
which is appropriate:
\debut
cycles &\Longleftrightarrow& L'(A),
\qquad ~~~~around\quad\nu\simeq 0\non\\
forms &\Longleftrightarrow& r'(a) \non 
\fin 
In other words, on quantum Riemann surfaces forms and cycles
are on an equal footing.

This fact can be formulated in more mathematical terms:

\proclaim Proposition.
For generic values of the coupling constant $\xi$, there exist
two skew symmetric polynomials $C(A_1,A_2)$ and $C^*(a_1,a_2)$
such that if we decompose them as follows~:
\debut
C(A_1,A_2) &=& \sum_k\({ L_k(A_1)M_k(A_2)-L_k(A_2)M_k(A_1)}\) \non\\
C^*(a_1,a_2) &=& \sum_j\({ s_j(a_1) r_j(a_2) - s_j(a_2) r_j(a_1)}\) \non
\fin 
then the ``period matrix " $P$ defined by:
\debut
P_{ij}= \pmatrix{ \vev{L_i,r_j} & \vev{L_i,s_j} \cr
	\vev{M_i,r_j} & \vev{M_i,s_j} \cr }_{ij} \label{QRie}
\fin
is a symplectic matrix.
\par

For the proof, see ref.\cite{sm2}. A quick comparison with the previous
sections shows that this is really the quantum analogue of the
Riemann bilinear identity.

\subsection{Null-vector equations.}
As explained in \cite{bbs2}, the occurence of null vectors leads to a set 
of differential equations for the form factors, or for the correlation functions.
As in the classical theory, they reflect the quantum equations of motion. 

In ref.\cite{bbs2} these equations were written in a fermionic
language. Here we will rewrite them in an alternative, but
equivalent, bosonic language. We will just quote the results.
Thus, let us introduce again
the generating function of the descendent operators:
\debut
\CL(t,y) = \exp\({\sum_{k\geq 1} t_{2k-1}I_{2k-1}
+ y_{2k} J_{2k} }\)~\cdot~ {\bf 1} \label{Lbis}
\fin
The functions $\CL(t,y)$ may be understood as the generating function
of the expectation values of the descendents of the 
identities between any states of the theory. Choosing these
states to be the $n$-soliton states allows us to identify
$\CL(t,y)$ with the generating function of the form factors. But 
choosing these states to be those created by auxiliary operators
allows us to interpret $\CL(t,y)$ as the generating function
of the correlation functions.

\proclaim Proposition.
The equations arising from the ``exact forms" can be written as~:
\debut
\int dD~ e^{-\xi(D,y)}~ \CL(t+[D]_o;y+[D]_e) = 0 
\label{Qquant}
\fin
where $\xi(D,y)=\sum_{k\geq 1} D^{2k} y_{2k}$ and
$[D]_o=(\cdots,\({\frac{1-q^{2k-1}}{1+q^{2k-1}} }\) \frac{D^{-2k+1}}{2k-1},\cdots)$
and $ [D]_e= (\cdots, \frac{D^{-2k}}{k},\cdots)$. \\
The equations arising from the ``deformed Riemann bilinear identities"
can be written as~:
\debut
\int\limits_{|D_2|>|D_1|} dD_1 dD_2 (D_1^2-D_2^2)\tau(\frac{D_1}{D_2})~
e^{-\xi(D_1,y)-\xi(D_2,y)}~ \CL(t+[D_1]_o+[D_2]_o;y+[D_1]_e+[D_2]_e)=0
\label{Cquant}
\fin
where the function $\tau(x)$ is defined by:
\debut
\tau(x)=\sum\limits _{k=1}^{\infty}{1-q^{2k-1}\over 1+q^{2k-1}}x^{2k-1} -
\sum\limits _{k=1}^{\infty}{1+q^{2k}\over 1-q^{2k}}x^{2k} \label{deftau}
\fin
\par
To make sense of the function $\tau(x)$ we have to assume that the 
parameter $q$ is not a root of unity.
The function $\xi(D,y)$ should not be confused with the
function $\zeta(A,t)$ which was introduced above.

Equations (\ref{Qquant},\ref{Cquant}) are linear equations
for the generating functions $\CL(t,y)$. Thus, they give linear relations
among the correlation functions of the descendent operators. They
do seem to give non trivial information on these
correlation functions until we find a way to close this hierarchy
of equations.

In ref.\cite{bbs2}, the null-vector equations (\ref{Qquant})
and (\ref{Cquant}) were used to show that the character of
the space of local fields obtained by the bootstrap
approach matches the character of the space of field
of the ultraviolet conformal field theory.

\section{A new description of the KdV hierarchy.}
The classical limit, which corresponds to $\nu\to\infty$,
of the quantum equations leads to a new
formulation of the KdV hierarchy. In this description
the fundamental variable is the generating function $S(A)$ of the
densities of the integrals of motion: 
\debut
S(A) = 1 + \sum_{k\geq 1} A^{-2k} S_{2k} 
= \exp\(-\sum_k \frac{A^{-2k}}{k} J_{2k} \) \non
\fin
To write the equations we introduce the generating function $\CL^{cl}(t,y)$
defined by~:
\debut
\CL^{cl}(t,y)= \exp\(\sum_{k \geq 1} y_{2k} J_{2k}(t) \) \non
\fin

\proclaim Proposition.
The KdV hierarchy then reduces in the set of two equations for $\CL^{cl}(t,y)$:
\debut   
\int D~ e^{-\xi(D,y)}~dI(D) \CL^{cl}(t,y+[D]_e) = 0 \label{newQcl}
\fin
and
\debut
&& \int\limits _{|D_2|>|D_1|} D_2D_1 (D_1^2-D_2^2)
\log\({1- \frac{D_1^2}{D_2^2}}\) ~e^{-\xi(D_1,y)-\xi(D_2,y)}
dI(D_1)dI(D_2) \CL^{cl}(t,y+[D_1]_e+[D_2]_e) +\non\\
&&~~~~~~~~~~~~~~~~~~~~~~ +
8\pi i\int dD D^3 ~ e^{-2\xi(D,y)}~ \CL^{cl}(t,y+2[D]_e) =0 \label{newCcl}
\fin
where $[D]_e= (\cdots, \frac{D^{-2k}}{k},\cdots)$ and
$\xi(D,y)=\sum_{k\geq 1} D^{2k} y_{2k}$ as before, and
$dI(D)= \sum_k D^{-2k}dD\d_{2k-1}$. 
The mixed operator $dI(D)$  acts on $\CL ^{cl}$ by 
differentiation with respect to the time variables, ie by $\partial _{2k-1}$.
\par

Equations (\ref{newQcl},\ref{newCcl}) provide a system of linear differential
equations for the Taylor coefficients of $\CL^{cl}(t,y)$. It
becomes a system of non-linear differential equations for the
$J_{2k}$ only after  the closure condition
$\CL^{cl}(t,y)= \exp\({\sum_{k \geq 1} y_{2k} J_{2k}(t) }\)$ has been
imposed. Ie.  we have to impose the following factorization relation~:
\debut
\CL^{cl}(t,y+[D]_e)=\inv{S(D)} \CL^{cl}(t,y) \label{closure}
\fin
One may think of this closure equation as a kind of Ward identity.
With this closure condition, eqs.(\ref{newQcl}) and (\ref{newCcl})
are completely equivalent to those of the KdV hierarchy.

\end{document}